\documentstyle[psfig]{mn}%

\def\sech{\mathop{\rm sech}\nolimits}

\def\pc{\,{\rm pc}}
\def\msun{\,{\rm M_{\odot}}}
\def\mmsun{\,{\rm mM_{\odot}}}
\def\kms{\,{\rm km s}^{-1}}

\def\kpc{\,{\rm kpc}}

\def\d{{\rm d}}

\def\i{\relax\ifmmode{\rm i}\else\char16\fi}
\def\e{{\rm e}}
\def\spose#1{\hbox to 0pt{#1\hss}}
\def\lta{\mathrel{\spose{\lower 3pt\hbox{$\mathchar"218$}}
     \raise 2.0pt\hbox{$\mathchar"13C$}}}
\def\gta{\mathrel{\spose{\lower 3pt\hbox{$\mathchar"218$}}
     \raise 2.0pt\hbox{$\mathchar"13E$}}}

\begin{document}

   \title[Cuspy dark-matter halos and the Galaxy]
	{Cuspy Dark-Matter Halos and the Galaxy}

   \author[Binney J.J. \& Evans N.W.]
          {J.J. Binney and N.W. Evans,\\
           Theoretical Physics, 1 Keble Road, Oxford OX1 3NP \\
           }
\date{Received ...; accepted ...}

\maketitle

\begin{abstract}
The microlensing optical depth to Baade's Window constrains the
minimum total mass in baryonic matter within the Solar circle to be
greater than $\sim 3.9 \times 10^{10}\msun$, assuming the inner Galaxy
is barred with viewing angle $\sim 20^\circ$.  From the kinematics of
solar neighbourhood stars, the local surface density of dark matter is
$\sim30\pm 15\msun\pc^{-2}$. We construct cuspy haloes normalised to
the local dark matter density and calculate the circular-speed curve
of the halo in the inner Galaxy. This is added in quadrature to the
rotation curve provided by the stellar and ISM discs, together with a
bar sufficiently massive so that the baryonic matter in the inner
Galaxy reproduces the microlensing optical depth. Such models violate
the observational constraint provided by the tangent-velocity data in
the inner Galaxy (typically at radii $2-4$ kpc). The high baryonic
contribution required by the microlensing is consistent with
implications from hydrodynamical modelling and the pattern speed of
the Galactic bar. We conclude that the cuspy haloes favoured by the
Cold Dark Matter cosmology (and its variants) are inconsistent with
the observational data on the Galaxy.
\end{abstract}


\section{Introduction} 

The Cold Dark Matter (CDM) cosmological model has recently enjoyed
great success on large scales. Observational data as diverse as
precision measurements of the cosmic background radiation (e.g., de
Bernardis et al. 2001; Halverson et al. 2001), measurements of
absorption lines in the spectra of distant quasars (e.g., Efstathiou,
Schaye \& Theuns 2000), measurements of the abundances of deuterium,
and other light elements (e.g., Burles, Nollett \& Turner 2001),
surveys of the positions in redshift space of hundreds of thousands of
galaxies (e.g., Peacock et al. 2001) and measurements of the
brightnesses of distant supernovae (e.g., Reiss et al. 1998;
Perlmutter et al. 1999) all point to a single cosmological model.  In
this model, the mean energy density takes the critical value that
makes space flat. Baryonic matter accounts for only a few percent of
this energy density, $\sim 70 \%$ of which takes the form of material
for which gravity acts repulsively. The remaining energy density
($\sim30\%$) is carried by collisionless massive particles of some
type.

The way in which gravity amplifies seed inhomogeneities in
collisionless dark matter (DM) to build up collapsed structures,
around which galaxies and clusters of galaxies assemble, has been
extensively studied over two decades. On large scales the predictions
of these studies yield the consistent interpretation of data alluded
to above. Hence, apparent conflicts with the data that have recently
emerged on sub-galactic scales have attracted a great deal of interest
(e.g., Spergel \& Steinhardt 2000; Sellwood \& Kosowsky 2000; Evans
2001). Many of these conflicts arise because CDM predicts that DM
halos have cuspy centres.

Only large and costly simulations of DM clustering are capable of
probing the structure of galactic halos on scales of a few
kiloparsecs, with the result that some controversy has surrounded the
proposal of Moore et al. (1998) that at the smallest radii the density
in a CDM halo becomes a power-law in radius, $\rho\sim r^{-\alpha}$
with $\alpha\sim1.4$.  Most protagonists now agree that agreement with
the simulation data can be achieved by fitting the two-power-law model
with
 \begin{equation}\label{moore}
\rho(r)\sim{ 1\over r^{\alpha}(r_s+r)^{3-\alpha}}.
\end{equation}
However, this functional form provides only an approximate description
of the data, in which the local logarithmic slope of the density
profile decreases continuously from $\sim3$ at large radii to zero at
the smallest radii.  Even in the largest simulations there are only a
few million particles in any given halo, so artificial discreteness
effects become important at some radius. At the smallest radius for
which discreteness effects are unimportant, the slope is steeper than
unity and may be as large as $1.4$.

Navarro, Frenk \& White (1997) observed that the CDM halos that formed
in many different cosmological models could be well represented by the
double-power-law model (1) with $\alpha=1$, and this is now known as
the NFW profile. If every halo conforms to the NFW profile, the halo
can be characterized by just two numbers: its peak circular speed
$v_{\rm max}$ (which occurs at $r=2.2r_s$) and the ratio $c\equiv
r_{200}/r_s$ of the scale radius to the radius within which the mean
density exceeds the cosmic density by a factor of 200.  In any given
simulation, $v_{\rm max}$ and $c$ are correlated in the sense that
halos with the largest concentrations $c$ have the smallest rotation
velocities. The quantitative details of the correlation differ from
one cosmological model to another. Although the numerical evidence is
weaker because only CDM clustering has been simulated at the highest
resolution, there are clear indications that cuspy halos are a generic
feature of DM clustering (Moore 1999; Knebe et al.  2001). 

Since we live in a galaxy that is prototypical of the galaxies that
together make the largest contribution to the cosmic luminosity
density, it is natural to ask whether the predictions of CDM
simulations agree with the data for the Milky Way. Navarro \&
Steinmetz (2000) and Eke, Navarro \& Steinmetz (2001) have addressed
this question by asking what values of $v_{\rm max}$ and $r_{200}$ one
should associate with the Milky Way, and then inferring from them the
appropriate value of $c$. Once these three parameters have been
chosen, the profile that the dark matter had prior to the infall of
the Galactic baryons is determined. The dark-matter mass predicted by
this profile interior to the solar radius $R_0$ should not be larger
than the current dark-matter mass inside $R_0$. Whether this condition
is in fact satisfied, is controversial.

Here we argue that the predictions of CDM simulations are more
directly and effectively tested by using measurements of
solar-neighbourhood dynamics to normalize the density of dark matter
at the Sun. With this normalization, the predicted density of dark
matter a few kiloparsecs from the Galactic Centre conflicts strongly
with astrophysical constraints for $\alpha\gta 1$. Unless the
logarithmic slope of the dark-matter density profile somehow decreased
as baryons fell into the Galactic dark halo, this constraint on
$\alpha$ strongly conflicts with all current DM simulations.

\section{The Local DM Density}

By measuring the line-of-sight velocities and distances of K dwarf
stars seen towards the south Galactic pole, Kuijken \& Gilmore (1991)
showed that at the solar radius $R_0$ there is
$\sim71\pm6\,\msun\pc^{-2}$ of material within $1.1\kpc$ of the
Galactic plane. This result has been confirmed in several independent
reworkings of the original data (e.g., Olling \& Merrifield 2001).
Measurements of the proper motions and parallaxes of stars that lie
within $200\pc$ of the Sun have yielded estimates of the local density
of all matter: $(76\pm15)\mmsun\pc^{-3}$ (Cr\'ez\'e et al. 1998);
$(111\pm10)\mmsun\pc^{-3}$ (Pham 1997); $(102\pm6)\mmsun\pc^{-3}$
(Holmberg \& Flynn 2000). Since a uniform sheet $2.2\kpc$ thick and
with a density of $100\mmsun\pc^{-3}$ would have a column density
three times larger than the $71\msun\pc^{-2}$ measured by Kuijken \&
Gilmore, it is clear that the Galactic mass density is strongly
flattened close to the plane, presumably because the Galactic disc
makes a large contribution.

By counting disk M dwarfs in {\it Hubble Space Telescope} fields,
Zheng et al. (2001) found that the vertical profile of these objects
is well modelled by
 \begin{equation}\label{GBFprof}
\nu(z)=0.435\sech^2(z/270\pc)+0.565\exp(-|z|/440\pc).
\end{equation}
Since the majority of the disc's stellar mass must be contained
in such faint red stars and other objects such as white dwarfs that
trace the integrated star-formation history of the disc, we may safely
assume that the overall stellar mass density shares this profile. From
equation (\ref{GBFprof}), we have that
\begin{equation}
\hat z\equiv{1\over\nu(0)}\int_{-1.1\kpc}^{1.1\kpc}\d z\,\nu(z)=691\pc.
\end{equation}
By counting stars within $5\pc$ of the Sun (which can be detected
through their large proper motions) and using Hipparcos parallaxes
Jahrei{\ss} \& Wielen (1997) find that stars contribute
$39\mmsun\pc^{-3}$ to the mass density at the plane. Multiplying this
density by the effective disc thickness $\hat z$, we have that stars
contribute $26.9\msun\pc^{-2}$ to the $71\pm6\msun\pc^{-2}$ of matter
that lies within $1.1\kpc$ of the plane. The contribution of the
interstellar medium to this figure is uncertain, both because the
densest gas is patchy, and because direct detection of H$_2$ is
difficult. Olling \& Merrifield (2001) conclude that when helium is
included, gas contributes $13.7\msun\pc^{-2}$.\footnote{It is
reassuring that the surface density from the ISM, averaged over a
layer $\sim300\pc$ thick, yields a density $\sim46\mmsun\pc^{-3}$ that
agrees with the difference between the dynamically determined volume
density ($76$ to $110\mmsun\pc^{-3}$) and the $39\mmsun\pc^{-3}$
contributed by stars.}  Thus, $\sim41\msun\pc^{-2}$ of the mass within
$1.1\kpc$ of the plane can be accounted for by baryons, and the
remaining $\sim30\msun\pc^{-2}$ should be contributed by particle
DM. The error on this last number is large and uncertain.  In addition
to the errors reported by Kuijken \& Gilmore ($\pm6\msun\pc^{-2}$),
there should be contributions from the work of Zheng et al. (2001),
Jahrei{\ss} \& Wielen and the sources cited by Olling \&
Merrifield. The overall error could easily be as large as
$15\msun\pc^{-2}$.

\begin{figure*}
\centerline{\psfig{file=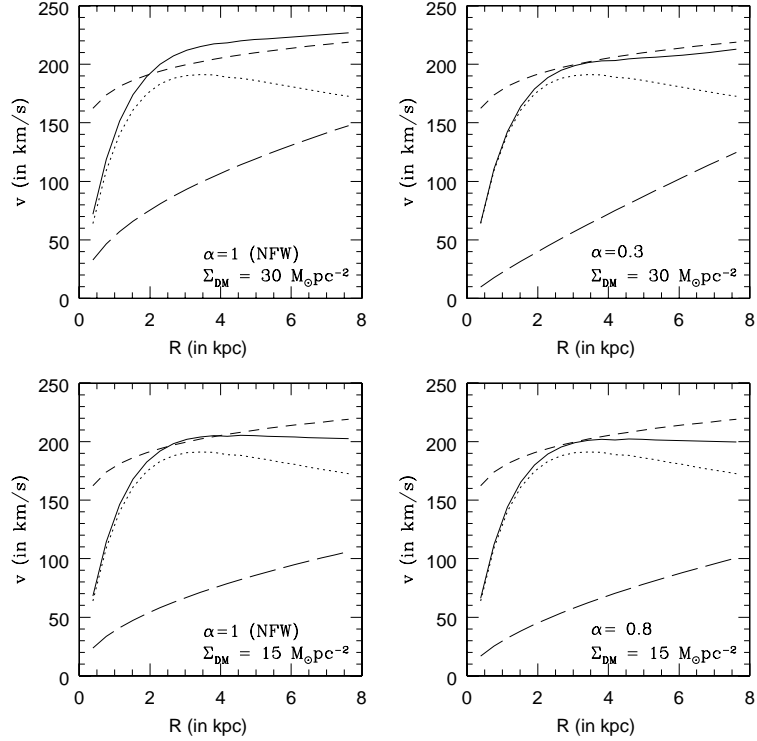,width=.6\hsize}}
\caption{The panels show the circular-speed curves generated by the
gas disc together with enough stars to yield $\tau = 2 \times 10^{-6}$
(dotted curves) and by cuspy haloes (long-dashed curves). The combined
rotation curve is shown as a solid line and must lie below the fit to the
tangent-velocity data (short-dashed line). The models differ in the
cusp index of the halo $\alpha$ and the local DM
contribution.\label{fig_one}}
\end{figure*}

\section{Microlensing constraints}

A crucial difference between baryonic and particle matter is that the
former can cause microlensing events, while the latter cannot. All
published estimates of the optical depth to microlensing of sources
near the Galactic Centre are nearly an order of magnitude greater than
the value $4 \times 10^{-7}$ that was originally anticipated
(Paczy\'nski 1991, Griest et al. 1991), so it is of interest to ask
what is the minimum baryonic mass in the inner Galaxy that can yield
the measured values.  This question is readily answered if one accepts
(i) that the density can never increase as one moves vertically away
from the Galactic plane, and (ii) that the Galaxy is
axisymmetric. Then the greatest optical depth per unit mass for
microlensing a source that lies distance $z_0$ from the centre along
the symmetry axis is obtained by forming the mass into a uniform
cylinder of any radius $R$ that just barely reaches up from the plane
to clip our line of sight to the star;\footnote{We assume that the Sun
lies in the plane; in reality it lies $\sim14\pc$ above the plane
(Binney, Gerhard \& Spergel 1997).}  the height $h$ of the optimal
cylinder is therefore
\begin{equation}\label{givesh}
h(R)=z_0(1-R/R_0).
\end{equation}
The mass $M$ and optical depth $\tau$ of this optimal cylinder are
related by
 \begin{equation}\label{MKuijken}
M={c^2z_0\over G}\tau,
\end{equation}
independent of $R$ (Kuijken 1997). In reality, the density of no
component can be constant for $z<h$ and then abruptly drop to zero. It
inevitably tapers smoothly to zero. If the vertical density profile
is either an exponential in $z$, with the scale height $h$ chosen to
maximize $\tau/M$, or a Gaussian in $z$ with the dispersion $h$ chosen
to maximize $\tau/M$, then $h$ is given by equation (\ref{givesh}) and
$M$ has to be larger than the value given by equation (\ref{MKuijken})
by a factor of either e or $(\pi\e/2)^{1/2}=2.07$, respectively. For
simplicity, we plot results based on exponential profiles (which fit
brightness profiles much better than do Gaussians). Hence, we assume
\begin{equation}\label{givesM}
M={\e c^2z_0\over G}\tau.
\end{equation}
Ours is a barred galaxy, and the optical depth of a given cylinder can
be increased by distorting it into a suitably aligned ellipse. Let
$\phi$ be the angle between the major axis of the ellipse and the Sun
-- centre line.  Then Binney, Bissantz \& Gerhard (2000) showed that
$\tau/M$ is maximized if the ellipse has axis ratio
\begin{equation}\label{givesq}
q=b/a=\tan\phi
\end{equation}
and that for this optimal axis ratio the mass required to achieve a
given optical depth is smaller than the value given by equation
(\ref{givesM}) by a factor
\begin{equation}\label{givesMq}
{M(q)\over M(1)}={1\over q}\sin^2\phi+q\cos^2\phi = 0.64, 
\qquad\quad (\phi=20\deg).
\end{equation}
The angle between the Galactic bar and the Sun--centre line is
believed to lie near $20\deg$ (Binney et al. 1991, 1997; Englmaier \&
Gerhard 1999), and we adopt this value in the following.

Following Binney et al. (2000), we ask what circular-speed curve is
implied if we structure the baryonic Galaxy so that (i) there is a
given optical depth to microlensing for a source that lies a distance
$z_0$ down the symmetry axis, (ii) the vertical density profile is
exponential in $z$ with the scale height given by equation
(\ref{givesh}) and (iii) the Galaxy inside $R_0/2$ is elliptical with
the axis ratio of equation (\ref{givesq}). The circular-speed curve we
derive should everywhere lie below the observational curve. Our
circular-speed curves differ from those presented by Binney et al.\ in
that here we include the contribution from the interstellar medium,
the radial profile of which we take to follow equation (1) of Dehnen
\& Binney (1998),\footnote{The first term in the exponential in this
equation should be $R_m/R$.} with the surface density scaled to
$13.7\msun\pc^{-2}$ at $R_0$.

The latest measurements of the optical depth to bulge sources both
come from the MACHO collaboration. From a Difference Imaging Analysis
(DIA) of the data, Alcock et al. (2000) estimate from 99 events
 \begin{equation}\label{DIAtau}
\tau=(3.2\pm0.5)\times10^{-6}\hbox{ at }(l,b)=(2.68\deg,-3.35\deg).
\end{equation}
From an analysis of 52 events in which a clump giant was lensed,
Popowski et al.\ (2000) find 
 \begin{equation}\label{Poptau}
\tau=(2.0\pm0.4)\times10^{-6}\hbox{ at }(l,b)=(3.9\deg,-3.8\deg).
\end{equation}
The first estimate is based on events in which faint stars are lensed,
so blending may render the optical depths uncertain.  Therefore, we
focus on the second estimate, which is based on the lensing of
relatively bright stars.  Moreover, the luminosity distribution of
these stars provides strong evidence that they really do lie close to
the Galactic Centre rather than far behind it, where they would have
larger optical depths. We may safely disregard the fact that the
optical depth has not been measured at $l=0$ since the value at $l=0$
is bound to be greater than any value at $|l|>0$. Binney et al.\ show
that the non-zero width of the distribution of the distances of the
lensed stars increases the measured mean optical depth by $\lta5\%$
and may safely be ignored. Hence, the formulae (\ref{givesM}) and
(\ref{givesMq}) may be applied with
$z_0=R_0\tan(3.8\deg)=531\pc$. This gives $\sim 3.9 \times 10^{10}
\msun$ as a lower limit for the baryonic mass within the solar circle.
We remark that (\ref{Poptau}) is the lowest estimate of the
microlensing optical depth towards the Galactic Centre published to
date, so our analysis errs on the conservative side.

The short-dashed curve in the panels of Fig.~\ref{fig_one} shows the
relation $v_c=(R/R_0)^{0.1}220\kms$, which for $R_0=8\kpc$ provides a
reasonable fit to the tangent-velocity data after correction for
non-circular velocities induced by the Galactic bar (Binney et
al.~1991).  The dotted curve in the panels of Fig.~\ref{fig_one} shows
the circular-speed curve we obtain for $\tau$ given by equation
(\ref{Poptau}) when the radial distribution of mass is determined by
assuming that in addition to the gas disc, there is an exponential
stellar disc with scale length $3\kpc$ and local surface density
$26.9\msun\pc^{-2}$, and an elliptical, exponential disc with
scale-length $R_b=1\kpc$ that contains the balance of the material
that is required to make up the optical depth. The main exponential disk is
elliptical at $R<R_0/2$. These parameter choices
are designed to place as much mass as possible in the elliptical inner
galaxy, where it generates the most optical depth per unit
contribution to the circular-speed curve. Binney et al. (2000) discuss
alternative radial distributions of stellar matter, and argue that for
the given optical depth a lower circular-speed curve can only be
obtained by placing more mass in the disc at $R_0$ than has been
measured to be present.

The solid curves in the panels of Fig.~\ref{fig_one} show the circular
speeds obtained by adding in contributions from a dark matter halo of
general form (\ref{moore}).  The long-dashed curves show the
circular-speed curve of the halo in isolation.  In the upper panels,
the halo contributes $30\msun\pc^{-2}$ within $1.1\kpc$ of the plane;
in the lower panels, it contributes $15\msun\pc^{-2}$, the minimum
value suggested by the analysis of Section 2. In the leftmost panels,
the halo is an NFW model ($\alpha =1$) with concentration zero (i.e.,
$\rho\sim r^{-1}$ at all radii). Although the full curve does not go
much above the observational limit shown by the dashed curve, it
cannot be treated as consistent with the data because the scale-height
distribution of the stellar disc has been so ruthlessly manipulated to
maximize the optical depth to a source $531\pc$ along the symmetry
axis. In the rightmost panels, the index $\alpha$ is chosen to be as
large as possible, consistent with the total circular speed falling
below the tangent-velocity data. If the local halo surface density is
$30\msun\pc^{-2}$ within $1.1\kpc$, then $\alpha \sim 0.3$ is needed
before the solid line drops to below the observational limit; the
corresponding value if the halo surface density is $15\msun\pc^{-2}$
is $\alpha \sim 0.8$.

In all the panels of Figure 1, the full curve lies below the
short-dashed curve at $R\sim500\pc$. This discrepancy arises because a
large scale height has been adopted for the stellar distribution at
these radii, in order to bring stars into the line of sight from the
Sun to the sources (eq.~\ref{givesh}). Infrared photometry implies
that the stellar distribution does not have such a large scale height,
and the line-of-sight velocities of gas in the `$x_2$ disc' at
$R\lta200\pc$ imply that the circular speed there agrees with the
dashed rather than the full curve (e.g.\ Binney 1999). The obvious way
to increase $v_c$ at $R\lta1\kpc$ without pushing it up at
$R\sim3\kpc$ is to move mass down into the plane at $R\lta1\kpc$, thus
diminishing the optical depth to the sources. In other words, there is
clear dynamical evidence that the stellar distribution is not optimal
from the point of view of generating optical depth per unit mass, and
only a still more massive stellar distribution than we have assumed
will generate $\tau=2\times10^{-6}$. Consequently, the circular-speed
curves in Fig.~\ref{fig_one} should be moved up at $R\sim R_0/2$ to
reflect the less than perfect efficiency with which the stars generate
optical depth, and there will be correspondingly less room for the
halo to contribute to the circular speed.

\section{Discussion and Conclusions}
 
How can this head-on collision between the enormously successful DM
theory and microlensing measurements be resolved? First it is worth
reassuring oneself that the microlensing data we have used are
compatible with {\it some\/} galaxy model, for the older results were
not (Binney et al.~2000). Near-infrared light should be a good tracer
of stellar mass (Rix \& Rieke 1993), so it is instructive
to construct Galaxy models in which mass follows the light
distribution that is inferred from near-IR photometry (Binney et
al.~1997; Freudenreich 1998; H\"afner et al.~2000; Bissantz \& Gerhard
2002). The main problem here is the effect of extinction by dust,
which even in the K-band is non-negligible. The best available models
contain a bar about $3\kpc$ long with an axis ratio about $3:1$ whose
major axis is inclined at about $20\deg$ to the Sun--centre line. When
{\it all\/} the dynamically permitted mass is put into such a model,
it generates an optical depth $\tau\sim1.4\times10^{-6}$ to a point
$500\pc$ along the minor axis (Bissantz \& Gerhard, 2002). Thus, a
plausible model of the Galaxy has an optical depth that lies within
$\sim1.5\sigma$ of the measured value (\ref{Poptau}) if the DM content
of the inner Galaxy is entirely negligible.  Prior to the advent of
high-resolution simulations, this result would have been taken as
confirmation that DM halos have homogeneous cores, as was long assumed
to be the case for no compelling reason. Now we must ask if there are
any corroborating pieces of evidence that DM makes a negligible
contribution to the mass of the inner Galaxy. There are two, namely the
pattern speed of the Galactic bar and hydrodynamical modelling of
longitude-velocity data.

DM, being dissipationless, rotates much more slowly than the Galaxy's
baryons and will not participate fully in the Galactic bar.  The
response of the DM to the bar is out of phase with it, and this
provides a drag force or dynamical friction.  When the DM and baryons
provide comparable mass,  the dynamical friction is at its most
severe. This is exactly the r\'egime in which CDM finds itself.  Both
analytic and numerical arguments agree that if there is substantial DM
present in the bar region, dynamical friction slows the rotation rate
of a bar on the timescale of a few rotation periods (Weinberg 1985,
Debattista \& Sellwood 2000).  So, a powerful argument that DM is
negligible in the bar region (the inner 3kpc) comes from measurements
of the pattern speed $\omega_b$ of the Galactic bar.  Dehnen (1999)
shows that the velocity-space distribution of stars near the Sun that
can be constructed from Hipparcos data yields a direct estimate of
$\omega_b$ as a multiple of the circular speed at the Sun, $\Omega_0$:
the natural interpretation of the data requires the Sun to lie just
outside the bar's outer Lindblad resonance, so that
$\omega_b=(1.95\pm0.15)\Omega_0$. Confirmation of the high pattern
speed comes form the hydrodynamical simulations of Englmaier \&
Gerhard (2000), who obtain a slightly higher value of $\omega_b$ by
requiring the $3\kpc$ expanding arm to lie just within corotation. We
conclude that the Galactic bar is a fast bar, and its corotation
radius is comparable with its size as inferred from COBE/DIRBE data ($
\sim 3.5\kpc$).

Studies of the flow of gas at $R\lta4\kpc$ probe the Galactic potential by
seeking matches between features in simulated longitude-velocity plots with
those observed in the spectral lines of HI, CO and other molecules (e.g.,
Englmaier \& Gerhard 1999). These features are associated with orbital
resonances, so the longitudes at which they occur depend on both $\omega_b$
and the underlying circular-speed curve. Since non-circular motions are
driven by the non-axisymmetric component of the Galactic potential, the
extent in velocity of features is a measure of how much the bar contributes
to the overall potential, and therefore constrains the contribution to
the potential of DM. Englmaier \& Gerhard (1999) find little DM within the
inner 3 kpc. The spiral arms outside corotation are probably affected by the
dark halo. However, even at the Solar circle, the halo contribution to the
radial force is only $\sim 20 \%$ in their models. Similarly, the stellar
dynamical models (e.g., H\"afner et al. 2000), in which there is no DM in the
central regions, reproduce essentially all of the kinematic data on the
velocity dispersion and streaming of stars in the Galactic bulge.

We are therefore driven to the conclusion that the high microlensing
optical depth is in excellent agreement with two other strong pieces
of evidence on the distribution of baryons and DM in the inner
Galaxy. By combining the microlensing optical depth with the DM
contribution estimated from local kinematics, we have shown that even
the least concentrated NFW model is clearly ruled out, as it violates
the constraint on the Galactic rotation curve.  In fact, if the DM
locally contributes $\sim30\msun\pc^{-2}$ within 1.1 kpc of the
Galactic plane, then no halo with a cusp steeper than $\sim 1/r^{0.3}$
is viable. 

Popowski et al.'s (2000) value of the optical depth $\tau$ is
preliminary, though entirely consistent with the earlier results.  We
can estimate the maximum value of $\tau$ compatible with an NFW halo
in the following way. Let us use models composed of scaled copies of
our inner Galaxy together with any of the NFW dark haloes occupying
the shaded region of Figure 4 of Eke et al. (2001). These are
$\Lambda$CDM haloes which satisfy the constraint that the dark matter
within the solar circle is less than $4.3 \times 10^{10} \msun$ and
are advocated by Eke et al. as satisfactory representations of the
Galactic halo.  We require that the combined circular-speed lie below
the relation $v_c=(R/R_0)^{0.1}220\kms$ (the short-dashed curve in
Figure 1). If the NFW halo has a local DM surface density
$\sim30\msun\pc^{-2}$, then the inner Galaxy must be scaled by $\sim
60 \%$, whilst if the surface density is $\sim15\msun\pc^{-2}$, then
the scale factor is $\sim 80 \%$. So, pushing all parameters to their
very extremes and assuming that the baryonic matter is optimally
configured in the inner Galaxy, then an NFW halo is compatible with an
optical depth $\tau \sim 1.6 \times 10^{-6}$.

\section*{Acknowledgments}
We thank Vincent Eke for kindly providing us with data from his
numerical simulations. NWE is supported by the Royal Society.

{}

\label{lastpage}
\end{document}